\documentstyle[12pt,aasms4]{article}

\received{1996 December 19}
\accepted{1997 February 6}
\journalid{ApJ Letters}{}
\articleid{480}{}

\lefthead{Chaplin et al.}
\righthead{The solar core: recent results from BiSON}

\begin{document}

\title{The solar core: new low-$\ell$ $p$-mode fine-spacing results
from BiSON\footnote{BiSON home page: http://bison.ph.bham.ac.uk}}

\author{W.~J.~Chaplin\footnote{E-mail: wjc@star.sr.bham.ac.uk},
Y.~Elsworth, G.~R.~Isaak, C.~P.~McLeod, B.~A.~Miller}

\affil{School of Physics \& Space Research, University of Birmingham,
Edgbaston, Birmingham B15 2TT, U.K.}

\and

\author{R.~New} \affil{School of Science \& Mathematics, Sheffield
Hallam University, Sheffield S1 1WB, U.K.}

\begin{abstract}

The fine-structure spacing $d_{\ell}(n) = \nu_{\ell,n} -
\nu_{\ell+2,n-1}$ for low-degree solar $p$ modes of angular degree
$\ell$ and radial order $n$ is sensitive to conditions in the deep
radiative interior of the Sun. Here we present fine-structure
spacings derived from the analysis of nearly five years of
helioseismological data collected between 1991 July and 1996 February
by the Birmingham Solar-Oscillations Network (BiSON).  These data
cover $9 \le n \le 28$ for $d_{0}(n)$, and $11 \le n \le 27$ for
$d_{1}(n)$.  The measured spacings are much more precise and cover a
greater range, than earlier measurements from BiSON data (Elsworth et
al. 1990a).  The predicted fine-structure spacings for a ``standard''
solar model are clearly excluded by the BiSON data (at $\approx
10\sigma$); models that include helium and heavy element settling
provide a much better match to the observed spacings (see also
Elsworth et al. 1995). Since the inclusion of core settling in solar
models will tend to slightly increase the predicted neutrino flux,
the BiSON fine-structure data appear to reinforce previous
conclusions, i.e., an astrophysical solution to the solar neutrino
problem seems unlikely. 

\end{abstract}

\keywords{Sun: interior -- Sun: oscillations -- Stars: evolution}

\section{INTRODUCTION}
\label{sec:intro}

The solar $p$-mode oscillations are manifestations of trapped
standing acoustic waves in the solar interior. The low-degree
(low-$\ell$) modes, which are globally coherent, possess radial wave
functions that penetrate into the core of the Sun. Fine-structure
spacings $d_{\ell}(n) = \nu_{\ell,n} - \nu_{\ell+2,n-1}$ (where $n$
is the modal radial order), constructed from the low-degree modes,
are extremely sensitive to conditions in the deep solar interior. The
use of these spacings to constrain models and theory is particularly
appealing, since one avoids relying upon inadequate modelling of the
outer solar layers.

Here we present fine-structure spacings $d_{0}(n)$ and $d_{1}(n)$
derived from the analysis of several years of solar Doppler velocity
data collected between 1991 and 1996 by the Birmingham
Solar-Oscillations Network (BiSON). Elsworth et al.  (1990a), with
pre-1990 vintage BiSON data, demonstrated that the ``standard'' solar
model provided a good match to the observed spacings. Further, a
variety of WIMP models (e.g., Cox, Guzik \& Raby 1990) were shown to
be in significant conflict with the observations.  Elsworth et al.
(1990a) therefore concluded that the helioseismological data provided
evidence for the solar neutrino problem requiring a solution in
particle -- and not astro -- physics, e.g., via mixing between
different species of neutrinos or spin precession.  The substantially
superior quality of the BiSON data presented in this paper has
allowed us to make even more precise -- and demanding -- comparisons
with solar models. We demonstrate that models including helium and
heavy element settling are in much better agreement with the new
BiSON data (see also Elsworth et al. 1995).

\section{DATA AND ANALYSIS}
\label{sec:data}

We have used seven 8 month-long, and one 32 month-long time series,
all constructed from BiSON data collected between 1991 July and 1996
February, for the analyses discussed in this paper.  The modes have
been fitted in the frequency domain as rotationally split, Lorentzian
multiplets with an associated flat background offset. The relative
amplitudes of different $m$ components within individual $\ell$
multiplets were fixed with the whole-disc sensitivities of
Christensen-Dalsgaard (1989).  We have constrained the heights of
peak components with the same $|m|$ to be equal within each fitted
multiplet.  A Levenberg-Marquardt technique was used to perform the
fitting, minimizing a maximum likelihood function that reflects the
$\chi^2$ 2-d.o.f. statistics of the frequency spectrum (Chaplin et
al. 1996). Dziembowski \& Goode (1996) indicate that $m$-dependent
asymmetries in the real mode-multiplet structures might introduce
systematic errors. Our mode-fitting procedure is most heavily
weighted toward the sectoral components, which appear as the
strongest peaks in our data. The calculations of these authors
suggest that -- considering, for example, $d_{0}(n)$ -- the frequency
differences between the $\ell=0$ and $\ell=2$, $m= \pm 2$ components
should not be strongly affected over the solar cycle.

Fine structure spacings $d_{0}(n)$ and $d_{1}(n)$ were determined for
each spectrum from the fitted frequencies. Formal errors on the
spacings were derived from the formal uncertainties on the mode
frequencies returned by the fitting program.  Mean sets of spacings
were constructed from the seven 8-month spectra by computing weighted
averages according to
 \begin{equation}
 \bar{d}_{\ell}(n) = \frac{
 		  \displaystyle\sum_{i=1}^{N} d_{\ell}(n)_{i}/\sigma_{i}^2}
   	          {\displaystyle\sum_{i=1}^{N} 1/\sigma_{i}^2},
 \label{eq:mean}
 \end{equation}
 where $N$ is the number of independent measures at each $n$, i.e.,
$N \le 7$, and $\sigma_{i}$ are the errors associated with each
spacing $d_{\ell}(n)_{i}$. Estimates of the external error on each
weighted mean (e.g., see Topping 1962) were calculated according to
 \begin{equation}
 \delta_{\rm ext} [\bar{d}_{\ell}(n)] = t[N-1] \cdot \sqrt{
 			 \frac{
	      \displaystyle\sum_{i=1}^{N} \frac{(d_{\ell}(n)_{i}-
			 \bar{d}_{\ell}(n))^2}{\sigma_{i}^2}}
	      {(N-1) \displaystyle\sum_{i=1}^{N} 1/\sigma_{i}^2},
			      }
 \label{eq:error} 
 \end{equation}
 where, because of the small number of samples at each ($\ell,n$), a
suitable correction factor, $t[N-1]$, appropriate to the desired
$1\sigma$ significance interval, has been drawn from the $t$
distribution in order to correctly assess the confidence limits
bracing the mean (e.g., see Book 1978, p. 85).  Constructing the
errors in the manner shown above reflects the scatter in the
observables, i.e., an analysis indicates that the external errors are
a factor of $\approx 1.5$ larger than the internal errors, while
preserving the additional information provided by the formal fitting
uncertainties.  A small number of data were omitted from a few of the
mean spacing calculations on the basis of several ``outlier''
rejection criteria.  We employed variants of the standard discordancy
test (internally studentized extreme deviation from the sample mean;
e.g., see Barnett \& Lewis 1984, p. 167), omitting each measure in
turn from the sample in the same spirit as the ``jackknife''
technique and also taking into account the internal weights in one of
the tests.  In addition, we also used the Dixon $Q$ family of tests
(Dixon 1951) in order to check for the presence of single (Dixon's
$r_{10}$) and double (Dixon's $r_{20}$ and $r_{21}$) outliers. 

Final sets of fine-structure spacings were then constructed by
combining the 32-month, and mean 8-month data: for $d_{0}(n)$,
32-month spacings were used for $9 \le n \le 13$, and mean 8-month
spacings for $14 \le n \le 28$; for $d_{1}(n)$, 32-month spacings
were used for $11 \le n \le 13$, and mean 8-month spacings for $14
\le n \le 27$. The errors used for $n \le 13$ are the formal
uncertainties from the fitted 32-month spectrum, while those above
are the external errors from the 8-month averages
(Equation~\ref{eq:error}). The averaged data sets can be found at
{\tt http://bison.ph.bham.ac.uk}.

\section{RESULTS AND DISCUSSION} 
\label{sec:models}

\begin{figure*}
 \epsscale{0.67}
 \centerline{\plotone{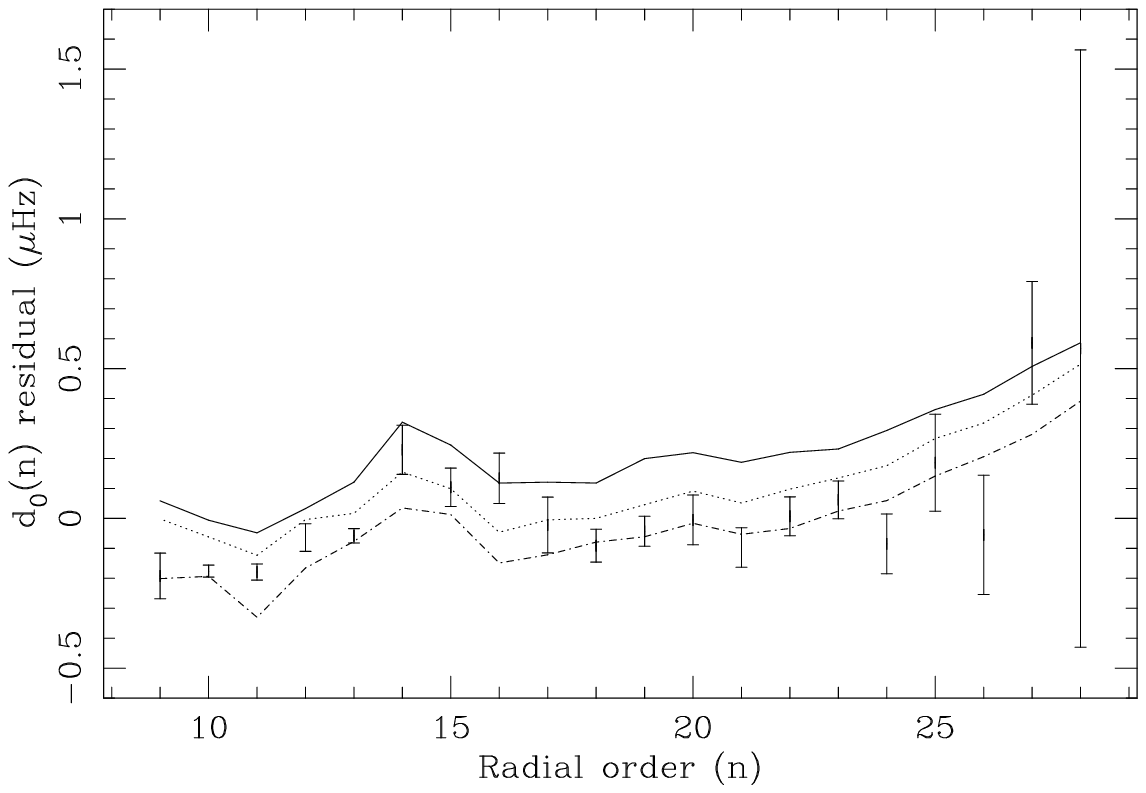}}
 \centerline{\plotone{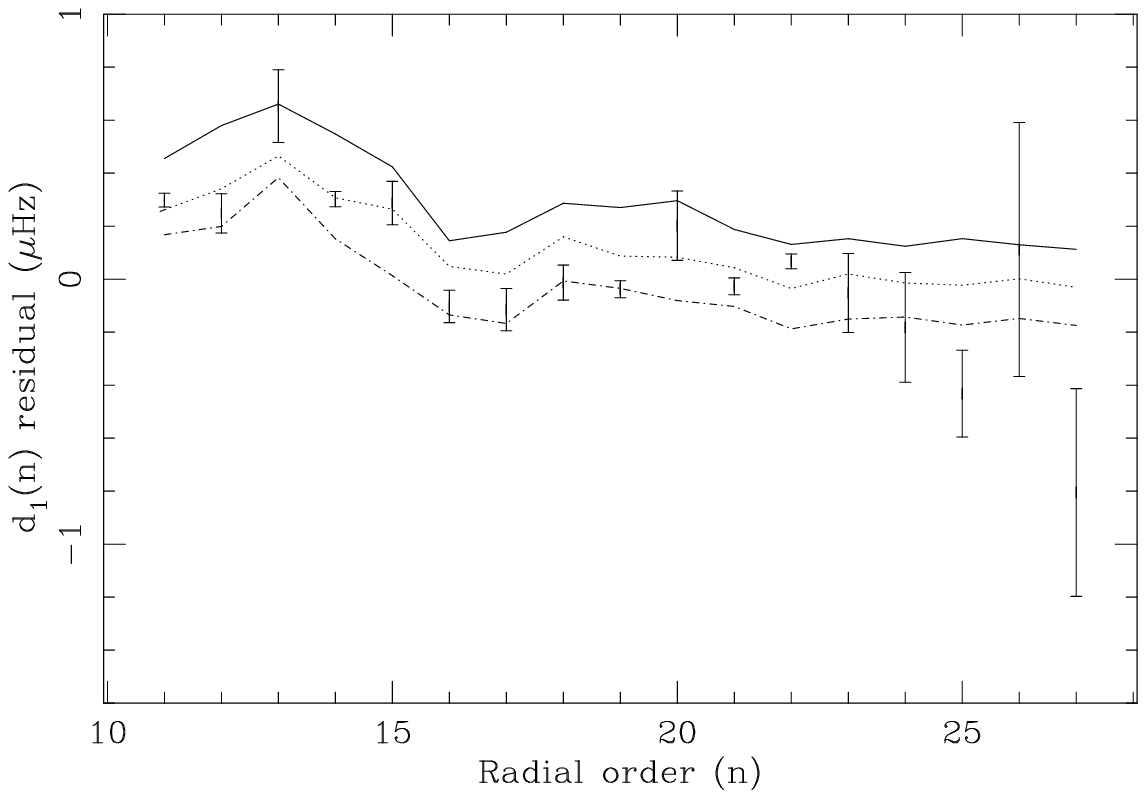}}
 \caption{The upper [$\bar{d}_{0}(n)$] and lower
 [$\bar{d}_{1}(n)$] plots show residuals generated by subtracting the
 appropriate best straight-line fit $c_{0}+c_{1} \cdot (n-21)$ from
 the mean fine-structure spacings. The solid lines on each plot show the
 residuals from a ``standard'' solar model (A), while the dot-dashed (B)
 and dotted (C) lines show the residuals from models which include
 helium and heavy element settling (see \S~\ref{sec:models}).}
 \label{fig:fineres}
 \end{figure*}

We have followed the convention of Elsworth et al.  (1990a), and
parameterized each set of data in terms of a straight line according
to
 \begin{equation}
 \bar{d}_{\ell}(n) = c_{0}+c_{1} \cdot (n-21),
 \label{eq:param}
 \end{equation}
 where $c_{0}$ and $c_{1}$ are the coefficients of the fit, suitably
normalized to the radial order datum $n=21$ on the abscissa. We have
performed the fit over the range $15 \le n \le 27$ in order to
facilitate a proper comparison with Elsworth et al. (1990a). The
best-fit coefficients for the Elsworth et al. (1990a) and new BiSON
data are given in Table~\ref{table:comp}. The new data are
represented graphically in Fig.~\ref{fig:fineres}: here, the
residuals, generated by subtracting the appropriate best
straight-line fit from the mean fine-structure spacings
$\bar{d}_{0}(n)$ and $\bar{d}_{1}(n)$, have been plotted.  (See later
for explanation and discussion of the solid, dot-dashed and
dotted-line model-generated predictions.)  The uncertainties on the
new best-fit values are, as expected, substantially superior to those
in Elsworth et al. (1990a), owing to the higher quality and quantity
of the more recent data. The fitted $c_{0}$ coefficients for
$d_{1}(n)$ differ by some $3\sigma$ (combined error): this is because
the $\ell=3$ frequencies -- the mode-peaks of which are substantially
weaker than those at $\ell=0$, 1 and 2 -- were less-well determined
in the old BiSON spectra.

Fitting statistics indicate that a straight line is an inadequate
representation of both sets of data, i.e., with reference to
Fig.~\ref{fig:fineres}, there is clearly rather more structure
present in each plot. The use of higher-order terms, or other
more-complicated functions, in any parameterized description of
$\bar{d}_{0}(n)$ and $\bar{d}_{1}(n)$ is therefore now required,
given the accuracy of the measured spacings.

We have searched for solar-cycle effects in both $d_{0}(n)$ and
$d_{1}(n)$.  No significant variations were found.  For $d_{0}(n)$,
we find that a variation in the mean spacing, as weighted by the
global view, over the range $14 \le n \le 26$, is excluded at $\sim
0.08\,\rm \mu Hz$ ($3\sigma$); and for $d_{1}(n)$, over the same
range in $n$, it is excluded at $\sim 0.09\,\rm \mu Hz$ ($3\sigma$).
These values should be compared with the corresponding $\sim
0.45\,\rm \mu Hz$ change observed in the eigenfrequencies themselves
(Elsworth et al.  1990b, 1994; Libbrecht \& Woodard 1991; Regulo et
al.  1994; see also Evans \& Roberts 1992).  From an asymptotic
description of the acoustic modes (e.g., Tassoul 1980) it can be
shown that (for $n >> \ell$) the fine spacings are proportional to
the gradient of the sound speed.  Since $n/\ell$ is quite small for
some of the observed low-degree modes that make up the presented
fine-structure spacings, care must be taken in interpreting these
data through the use of asymptotics.  This assumed, the solar-cycle
constraints on $d_{0}(n)$ and $d_{1}(n)$ imposed by the BiSON data --
in each case an upper limit to any change of less than 1 per cent --
would appear to place similar tight constraints on any implied
variation of the sound-speed gradient through the deep solar interior
over the solar cycle. 

We have compared the observed BiSON fine-structure spacings with
those derived from three, slightly different solar models. These are
referred to as models A, B and C in Table~\ref{table:comp}.  Model~A
is a ``standard'' solar model (model~1 from Christensen-Dalsgaard,
Proffitt \& Thompson 1993; official designation 14b.14), which
neglects settling effects, and was constructed with the CEFF equation
of state (Christensen-Dalsgaard \& D\"appen 1992) and the Livermore
(OPAL) opacity tables (Iglesias, Rogers \& Wilson 1992).  Model~B
(model~4 from Christensen-Dalsgaard, Proffitt \& Thompson; official
designation 14b.d.20) again employs the CEFF equation of state and
the OPAL opacities, in addition to helium and heavy-element settling
with turbulent mixing.  Model~C (Reference model `S' from
Christensen-Dalsgaard, D\"appen et al.  (1996), which appears as
model OPAL~1 in Basu et al. (1996); official designation 15bi.d.15)
incorporates settling effects -- neglecting turbulence -- and employs
the OPAL opacities, in addition to the OPAL equation of state
(Rogers, Swensen \& Iglesias 1996). 

The fine-structure data for each model were fitted, as per
equation~\ref{eq:param}, over the range $15 \le n \le 27$. An
inspection of the fitted $c_{0}$ coefficients listed in
Table~\ref{table:comp} clearly indicates that model~A is excluded at
a high level of significance. While models B and C provide a better
match to the observed data, the $d_{0}(n)$ comparison favours model
B, while that for $d_{1}(n)$ favours model C. The fitted slopes for
the models are all fairly consistent: they lie roughly 3 and
$1.5\sigma$ from the new BiSON values for $d_{0}(n)$ and $d_{1}(n)$
respectively.

The use of the fitted BiSON and model coefficients in assessing the
relative merits of the models is, to some extent, misleading, i.e.,
the comparison is ultimately dependent upon the suitability of the
parameterized description -- here, a simple straight line -- used to
describe the trend in the spacings.  As previously noted, the plots
in Fig.~\ref{fig:fineres} clearly show that a straight-line
representation of the measured spacings is inadequate.  We have
therefore also employed a direct one-to-one comparison between the
BiSON and model spacings, as indicated in Table~\ref{table:comp}. For
each model under consideration, fine-spacing residuals were computed
at each $n$, in the sense of BiSON minus model. Mean, weighted
residuals were then computed for a variety of ranges in $n$ -- those
for the range $9 \le n \le 28$ for $d_{0}(n)$, and $11 \le n \le 27$
for $d_{1}(n)$, are shown in Table~\ref{table:comp}.  The
uncertainties on each fine-structure datum were used to weight the
computation of the mean, and its error determined from the weighted
scatter of the residuals.

With reference to Table~\ref{table:comp}, the difference residuals
imply that the spacings for the ``standard'' model (A) are
significantly larger than the BiSON values. The mean residuals for
$d_{0}(n)$ and $d_{1}(n)$ exclude model A at the 13 and $\sim
9\sigma$ levels.  The models incorporating settling provide a much
better match to the observed spacings; however, marginally
significant differences do remain.  The comparative $d_{0}(n)$
statistics imply that model~B is in slightly better agreement with
the BiSON spacings than model~C.  Fig.~\ref{fig:fineres} appears to
show that model B (dot-dashed line) provides a better match to the
the BiSON $\bar{d}_{0}(n)$ data for $n > 17$, while at lower $n$
models B and C are comparable. (The weighted difference residuals for
$9 \le n \le 17$ are significant at the $\sim 3$ and $\sim 4\sigma$
levels for models B and C respectively; while for $18 \le n \le 28$
they are significant at the $\sim 0.1$ and $\sim 5\sigma$ levels
respectively.) For $\bar{d}_{1}(n)$, the difference residuals for
models B and C are similar; model C is perhaps more in accordance
with the observed values.

As an aside, we note that the inclusion of turbulence in model B has
little effect on the spacings. The introduction of turbulence in the
model changes the sound-speed profile, between approximately $0.5 \le
r/R_{\odot} \le 0.7$, relative to a model that neglects its effects.
However, the sound-speed profile -- and the hydrogen and helium
abundances -- are very similar in the deep radiative interior.
Christensen-Dalsgaard, Proffitt \& Thompson also constructed a
settling model that neglected turbulent effects: this model produces
$c_{0}$, $c_{1}$, and mean residual values in good agreement with
model~B. 

The plots in Fig.~\ref{fig:fineres}, and the comparative residual
statistics, show clearly that model A is excluded -- at a very high
level of significance -- by the BiSON data. While neither of the
``settling'' models considered provides a satisfactory match over the
whole range in $n$ for which the observed spacings are available,
they do nevertheless match the BiSON data far better.  The inclusion
in solar models of non standard settling effects in the core will
tend to raise the opacity of the deep radiative interior and increase
the solar neutrino flux, albeit rather modestly
(Christensen-Dalsgaard 1996).  The results in this paper would
therefore appear to reinforce the conclusions of Elsworth et al.
(1990a), in arguing against an astrophysical solution to the solar
neutrino problem. 

Christensen-Dalsgaard (1996) does, however, point out that certain
models can be constructed (e.g., Antia \& Chitre 1995) that match
observed solar neutrino capture rates and the helioseismological
data, but at the cost of employing what appear to be unrealistic
assumptions regarding the physics of the solar interior, for example,
models that arbitrarily ``juggle'' the comparative contributions of a
reduced core opacity (which will tend to reduce the fine-structure
spacings) and mixing in the core (which will tend to increase them).
At present there appears to be little justification for constructing
such models. We add an additional note of caution by indicating that
the characteristics of the $p$ modes in themselves -- which depend
upon the mechanical properties of the solar interior -- cannot
uniquely define the internal solar temperature and, by implication,
the expected neutrino flux.  Nevertheless, helioseismological data,
such as those presented in this paper, reinforce our previous
conclusions, i.e., an astrophysical solution to the solar neutrino
problem seems to be excluded. If the various nuclear cross sections
are correct, then electron neutrinos would therefore seem to
disappear between the solar core and terrestrial detectors.  Possible
causes are: oscillations between various species, or the precession
of a magnetic moment, both of which require neutrinos to have mass
(e.g., see Bahcall 1988).

\acknowledgments

We would like to thank all those who are, or have been, associated
with the BiSON global network. In particular: H.~B.~van~der~Raay,
H.~Williams, J.~Litherland and R.~Lines in Birmingham and P.~Fourie
at SAAO: and our hosts R.~Stobie (SAAO); the Carnegie Institute of
Washington; the Australia Telescope National Facility (CSIRO);
E.~J.~Rhodes (Mt.  Wilson, California); and members (past and
present) of the IAC, Tenerife. We are indebted to
J.~Christensen-Dalsgaard for providing the model-calculated
frequencies.  Some of the research presented above made use of the
Birmingham node of {\sc starlink}.  The 10.7-cm radio flux data were
obtained from the World Data Centre.  BiSON is funded by the UK
Particle Physics and Astronomy Research Council.

%%%%%%%%%%%%%%%%%%%%%%%%%%%%%%%%%%%%%%%%%%%%%%%%%%%%%%%%%%%%%%%%%%%%%%%%%%%

 \begin{deluxetable}{cccc}
 \footnotesize
 \tablecaption{Comparison between BiSON and model fine-structure data}
 \tablewidth{0pt}
 \tablehead{
 \colhead{Model}& 
 \colhead{$c_{0}$ ($\rm \mu Hz$)\tablenotemark{a}}&
 \colhead{$c_{1}$ ($\rm \mu Hz$)\tablenotemark{a}}& 
 \colhead{Mean residual ($\rm \mu Hz$)\tablenotemark{b}}
 }
 \startdata
 \cutinhead{$d_{0}(n)$}
 Elsworth et al.& $9.00 \pm 0.06$& $-0.290 \pm 0.030$& \phn...\nl
 Data, this paper& $9.017 \pm 0.023$& $-0.3255 \pm 0.0073$& \phn...\nl
 A\tablenotemark{1}& $9.266 \pm 0.019$& $-0.3001 \pm 0.0050$&
$-0.170 \pm 0.013$\nl
 B\tablenotemark{2}& $9.033 \pm 0.019$& $-0.2980 \pm 0.0053$&
$\phn0.050 \pm 0.016$\nl
 C\tablenotemark{3}& $9.143 \pm 0.019$& $-0.2953 \pm 0.0051$&
$-0.081 \pm 0.013$\nl
  \cutinhead{$d_{1}(n)$}
 Elsworth et al.& $15.60 \pm 0.10$& $-0.460 \pm 0.060$&
\phn...\nl
 Data, this paper& $15.903 \pm 0.017$& $-0.4004 \pm 0.0071$&
\phn...\nl
 A\tablenotemark{1}& $16.103 \pm 0.020$& $-0.4165 \pm 0.0053$&
$-0.201 \pm 0.023$\nl
 B\tablenotemark{2}& $15.789 \pm 0.015$& $-0.4112 \pm 0.0041$&
$\phn0.120 \pm 0.024$\nl
 C\tablenotemark{3}& $15.951 \pm 0.017$& $-0.4166 \pm 0.0045$&
$-0.019 \pm 0.024$\nl
 \enddata
 \tablenotetext{a}{Fine-structure data fitted to equation
 $\bar{d}_{\ell}(n) = c_{0}+c_{1} \cdot (n-21)$, over range $15 \le n \le
 27$ in order to facilitate comparison with Elsworth et al. (1990a).}
 \tablenotetext{b}{Weighted mean difference residual in sense BiSON
 minus model data. For $d_{0}(n)$, mean calculated over
 range $9 \le n \le 28$; for $d_{1}(n)$, over range $11 \le n \le
 27$.}
 \tablenotetext{1}{A ``standard'' solar model, with no settling effects.
 This is model~1 from Christensen-Dalsgaard, Proffitt \& Thompson
 (1993).}
 \tablenotetext{2}{Model~4 from Christensen-Dalsgaard, Proffitt \&
 Thompson, incorporating settling with turbulent mixing.}
 \tablenotetext{3}{Reference model `S' from Christensen-Dalsgaard,
 D\"appen et al. (1996), which incorporates settling effects, but
 neglects turbulence.}
 \label{table:comp}
 \end{deluxetable}

%%%%%%%%%%%%%%%%%%%%%%%%%%%%%%%%%%%%%%%%%%%%%%%%%%%%%%%%%%%%%%%%%%%%%%%%%%%%

% FIGURE: first version that does NOT print out the encapsulated
%         postscript files.

 \clearpage

% \figcaption[fine1a.ps,fine1b.ps]{The upper [$\bar{d}_{0}(n)$] and
% lower [$\bar{d}_{1}(n)$] plots show residuals generated by
% subtracting the appropriate best straight-line fit $c_{0}+c_{1}
% \cdot (n-21)$ from the mean fine-structure spacings.
% The solid lines on
% each plot show the residuals from a ``standard'' solar model (A),
% while the dot-dashed (B) and dotted (C) lines show the residuals from
% models which include helium and heavy element settling (see
% \S~\ref{sec:models}).  \label{fig:fineres}}

%
% Now the version that does...
%

%%%%%%%%%%%%%%%%%%%%%%%%%%%%%%%%%%%%%%%%%%%%%%%%%%%%%%%%%%%%%%%%%%%%%%%%

\end{document}